\documentclass[amsmath,amssymb,aps,prb,preprint,groupedaddress]{revtex4-1}
\usepackage{bm}       
\usepackage{graphicx} 
\usepackage{hyperref} 

\begin{document}


\title{Thermal Properties of Rung Disordered Two-leg Quantum Spin Ladders: Quantum Monte Carlo Study}


\author{Ulvi Kanbur}
\email[e-mail: ]{ulvikanbur@karabuk.edu.tr}
\affiliation{The Graduate School of Natural and Applied Sciences, Dokuz Eyl\"{u}l University, 35390  {I}zmir, Turkey}
\affiliation{Department of Physics, Karab\"{u}k University, 78050 Karab\"{u}k, Turkey}

\author{Hamza Polat}
\author{Erol Vatansever}
\affiliation{Department of Physics, Dokuz Eyl\"{u}l University, 35390 Izmir, Turkey}
\date{\today}

\begin{abstract}

A two-leg quenched random bond disordered antiferromagnetic spin$-1/2$ Heisenberg ladder system is investigated by means of 
stochastic series expansion (SSE) quantum Monte Carlo (QMC) method. Thermal properties of the uniform and staggered susceptibilities, the structure factor, the specific heat and the 
spin gap are calculated over a large number of random realizations in a wide range of 
disorder strength. According to our QMC simulation results, the considered system has a special 
temperature point at which the specific heat take the same value regardless of the 
strength of the disorder. Moreover, the uniform susceptibility is shown to display the same 
character except for a small difference in the location of the 
special point. Finally, the spin gap values are found to decrease with increasing disorder parameter 
and the smallest gap value found in this study is well above the weak 
coupling limit of the clean case.
\end{abstract}


\maketitle

\section{\label{sec:sec1}Introduction}
The effect of quenched randomness on zero- and finite-temperature properties of the different types of statistical model systems is 
one of the fundamental problems in condensed-matter physics. The spin-1/2 Heisenberg spin chains with disorder\cite{shuPropertiesRandomsingletPhase2016,shirokaOrderRandomnessOnset2019},  the spin-1/2 $J-Q$ model on a two dimensional (2D) 
square lattice \cite{liuRandomSingletPhaseDisordered2018}, and quantum spin chains with power-law long-range 
antiferromagnetic (AFM) couplings \cite{moureDisorderedQuantumSpin2018} are some of 
the recent model systems including quenched randomness. Low dimensional spin systems have also 
been an attractive topic of research thanks to the 
development of theoretical, experimental, and computational methods\cite{akimitsu_towards_2019,konig_renormalization_2018-1,patel_magnetic_2016,suzuki_first-principles_2015,yu_structural_2020,zhang_magnetic_2019,zhang_pressure-driven_2017,zhang_sequential_2018,vasiliev_milestones_2018}. Besides, most of the unique properties of high-$T_C$ superconductivity in cuprates are likely 
linked to the low dimensional systems. Among the low dimensional systems, quantum spin systems with AFM interactions show rich physical properties even in one dimension. 
For example, Haldane's conjecture states that AFM spin chains with integer spins exhibit a gapped spectrum 
which has been supported by theoretical \cite{affleck_proof_1986}, experimental 
\cite{buyers_experimental_1986} and numerical (QMC) \cite{nightingale_gap_1986} studies. Also, the spin-$1/2$ Heisenberg
coupled chains with an even number of legs have a finite spin gap ($\Delta$) to the lowest triplet excitation. 
Some ladder systems have an exponentially decaying the spin-spin correlation function and the uniform 
susceptibility, and this can be regarded as existence of a spin gap \cite{dagotto_surprises_1996}. 
The spin gap can be extracted from the uniform susceptibility $(\chi_u(T))$ which has the following form for the temperature regime $T<<\Delta$ \cite{sandvik_computational_2010,frischmuth_susceptibility_1996,white_equivalence_1996}, 
\begin{equation}
 \chi_u(T)\sim T^{-1/2}\mathrm{e}^{-\Delta/T}.
    \label{eq:susc}
\end{equation}

The width of the spin gap for a two-leg ladder spin system with the isotropic coupling constant $J$ can be roughly estimated 
as  $ \Delta \approx0.5J$ by using QMC techniques within a reasonable computational time on modern computers \cite{barnes_excitation_1993}. The value of the spin gap can be altered or even eliminated by including disorder in the spin-spin coupling \cite{melin_strongly_2002}, choosing a different kind of a lattice topology \cite{maiti_frustrated_2018,metavitsiadis_competing_2017} or appliying external magnetic fields \cite{chitra_critical_1997}. It has been shown that 
the spin gap is drastically reduced by a light doping on the pure system with non-magnetic impurities \cite{nagaosa_nonmagnetic_1996,korenblit_lightly_1997,mikeska_spin_1997}. The alteration of the spin gap due to the external effects has been investigated by also QMC simulations\cite{iino_effects_1996,motome_impurity_1996,lavarelo_magnetic_2013}. For instance, a quantum phase transition is observed in a two-leg ladder spin system with non-magnetic impurities\cite{motome_impurity_1996}. It is found that the random depletion of spins introduces a random Berry phase term into the 
nonlinear $\sigma$ model\cite{nagaosa_nonmagnetic_1996}. Besides, 
the magnetic field has some remarkable effects on the physical behavior of spin 
ladder systems\cite{giamarchi_coupled_1999,chitra_critical_1997,ruegg_thermodynamics_2008,wessel_field-induced_2001}. 

Various properties  of a wide range of different ladder models have been studied, like spin ladder systems with dimerization \cite{cabra_magnetization_2000,japaridze_magnetic_2009,chen_identification_2012,chen_ground_2014,kariyado_topological_2015,jahangiri_thermodynamic_2017,shahri_naseri_magnetic_2017}, zig-zag ladders \cite{chen_phase_2001,hoyos_phase_2004,bunder_generic_2009,danilovich_spin-singlet_2017,maiti_frustrated_2018}, mixed ladders\cite{kolezhuk_mixed_1998,batchelor_exactly_2001,aristov_ferrimagnetic_2004,batchelor_quantum_2004,zad_phase_2018}. 
A gapless phase has been found in two-leg zig-zag ladders with frustration by benefiting from  
exact diagonalization and density matrix renormalization group (DMRG) methods \cite{maiti_frustrated_2018}. 
A ferrimagnetic spin-1 and spin-1/2 mixed spin ladder  has been analyzed by 
using spin-wave theory and  bosonization techniques \cite{aristov_ferrimagnetic_2004}. Thermal and ground 
state properties of the similar ladder systems have been also studied by appliying QMC methods \cite{chen_ground_2014,
jahangiri_thermodynamic_2017,danilovich_spin-singlet_2017}. The presence of quenched bond randomness may significantly affect 
the thermal and magnetic properties of the considered system even at low disorder concentration values. Weakly disordered anisotropic 
spin-1/2 ladders have been handled perturbatively and different phases of the clean case have been detected as 
sensitive and insensitive to the changing disorder \cite{orignac_weakly_1998}.    Furthermore, critical properties of 
strongly disordered systems  have  been mainly studied with strong disorder renormalization group 
method \cite{igloi_strong_2005,vojta_phases_2013}, and in combination with the DMRG method \cite{melin_strongly_2002}.

Modern QMC techniques are powerful tools to study the disordered spin ladder systems. For instance,  the SSE QMC technique has been used to investigate the spin-1/2 Heisenberg quenched bond disordered ladders, and it has been  found that the neighbouring bond energies change sensitively with the position of the disorder in the spin-spin coupling term \cite{trinh_bond_2013}. In Ref.~\onlinecite{hormann_dynamic_2018}, some unusual and interesting effects of disorder on collective 
excitations have been reported with the calculation of the ground-state dynamic structure factor for a ladder system with 
bond disorder along the legs and rungs of the ladder. To the best of our knowledge, 
the static properties of such a  disordered model have not been investigated so far. In this paper, we investigate
the thermodynamic properties of a two-leg quantum spin ladder system including a quenched  bond 
randomness along only the rung direction. For this aim, we used the SSE QMC method with operator loop update  \cite{sandvik_stochastic_1999,sandvik_computational_2010} for 
varying values of the system parameters. In a nutshell, our QMC 
simulations shows that  the spin gap value tends to  decrease with an increment in the disorder ratio. 
Moreover, a crossing point has been detected at which the disorder ratio does not play a 
critical role on the numerical values of both the specific heat and the uniform 
susceptibility curves.

The rest of the paper is planned as follows: In Section \ref{sec:sec2}, we give the details of the model and the 
simulation method with a common notation and formalism. The numerical results and discussion are 
given in  Section \ref{sec:sec3}.
Finally, section \ref{sec:sec4} contains a summary of our conclusions.

\section{\label{sec:sec2}Model and Method}
\begin{figure}[ht!]
\includegraphics[scale=.85]{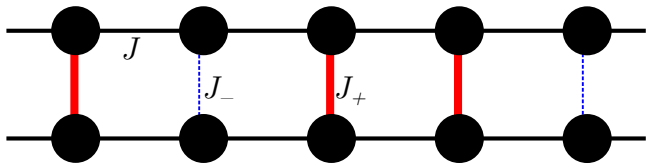}%
 \caption{\label{bimodal}(Color online)  An example of the quenched bond disorder configuration on a two-leg 
 ladder system. Solid (red) and dashed (blue) lines along the rungs are the bond couplings $J_+$ and $J_-$ 
 respectively, satisfying the condition $(J_++J_-)/2J=1.$ All the couplings ($J$) along the legs are the same.}
\end{figure}

We write the Hamiltonian of the quantum spin ladder model in a general manner to be in 
accord with the formulation of the SSE technique for convenience. The following Hamiltonian
\begin{equation}
 \mathcal H=\sum_b^{N_b}J_b\bm{S_{i(b)}\cdot S_{j(b)}}
\end{equation}
can technically describe a wide range of models consisting of $N_b$ bonds where a bond is a connected 
two sites ($i$ and $j$) with coupling strength $J_b$. Here, $\bm{S_{i(b)}}$ are spin operators at 
sites $i(b)$. For the present two-leg ladder model with $N$ sites, the bonds are all the nearest 
neighbour sites with $J_b>0$. The first $N$ bonds are along the legs with $J_b=J$, and the 
remaining $N/2$ bonds are along the rung direction with $J_b=J_+$ or $J_b=J_-$ that are selected 
randomly from a uniform distribution with equal probabilities, and they satisfy the condition $(J_++J_-)/2J=1$. An example 
of the quenched bond disorder on the system is shown in Fig.~\ref{bimodal}. Namely, the bonds along the rungs 
are drawn from the bimodal distribution 
\begin{equation}
\mathcal P(J_b)=p\delta (J_b-J_+)+(1-p)\delta(J_b-J_-)
\end{equation}
with probability $p=1/2.$

For the $S=1/2$ isotropic Heisenberg antiferromagnets ($J_b>0$) within the formulation of SSE technique, the bond 
operator $H_b=\bm{S_{i(b)}\cdot S_{j(b)}}$ can be divided into its diagonal and off-diagonal parts as follows, 
\begin{subequations}
\begin{equation}
H_{1,b}=\left(\frac14-S^z_{i(b)}S^z_{j(b)}\right) 
\end{equation}
\begin{equation}
H_{2,b}=\frac12\left(S^+_{i(b)}S^-_{j(b)}+S^-_{i(b)}S^+_{j(b)}\right)
\end{equation}
\end{subequations}
where $H_{a,b}$ is a diagonal and off-diagonal operator  for $a=1$ and $a=2$, respectively.
The Hamiltonian can be then rewritten as follows,
\begin{equation}
\label{eq:hamil}
 \mathcal H=-\sum_b^{N_b}J_b\left(H_{1,b}-H_{2,b}\right)+\mathrm{const.}
\end{equation}
where the constant energy term is not necessary for the implementation of the algorithm 
(but it should be added when calculating the energy). The non-zero matrix elements of the operators $H_{a,b}$ are all 
equal to $J_b/2.$ Concisely, SSE QMC technique, based on the Taylor series expansion of the partition 
function, can be formulated as a sum of the products of the operators $H_{a,b}$ with a fixed length scheme. 
More details including also the implementation of the algorithm can be found in Refs. \cite{sandvik_generalization_1992,sandvik_stochastic_1999,sandvik_computational_2010}. As a result, 
the full partition function can be  given as follows,
\begin{equation}
 \label{eq:full_part}
 \mathcal Z=\sum_{\alpha,S_L}(-1)^{n_2}\beta^n\frac{(L-n)!}{L!}\left<\alpha\left|\prod_{p=0}^{L-1}J_{b(p)}H_{a(p),b(p)}\right|\alpha\right>
\end{equation}
where the sums are over the configurations $\alpha$ and all possible operator products $H_{a,b}$ 
including additional unit operator $H_{0,0}$ and a coupling constant $J_0\equiv1$, on a string of 
length $L.$ Here $n$ and $n_2$ are the number of non-unit and off-diagonal operators on the string, 
respectively. $\beta$ is the inverse temperature with a unit Boltzmann constant $k_B$.  
The non-zero weights are bond dependent for an allowed 
configuration, and which can be written as follows
\begin{equation}
 \label{eq:weight}
 W(\alpha, S_{L})=\left(\frac{\beta}{2}\right)^{n}\frac{(L-n)!}{L!}\prod_{p=0}^{L-1}J_{b(p)}.
\end{equation}

The numerical results are obtained for the quenched random bond two-leg ladder system of  
the dimension $L_x\times2$. Here, $L_x=256$ is the system size along the legs of the ladder. For convenience, 
we define a disorder strength $\rho$ as $J_\pm=1\pm\rho$ where $J_+>1$ and $J_-<1$ for all 
values of $0\le\rho\le1$. $\rho=0.0$ corresponds to the clean case of the system. 

The specific heat $(\mathcal C)$ of the system can be easily measured by monitoring the number of 
non-unit operators $n$ in the operator sequence \cite{sandvik_computational_2010},
\begin{equation}
 \label{eq:specheat}
 \mathcal C=\left<n^2\right>-\left<n\right>^2-\left<n\right>.
\end{equation}
Static susceptibilities can be evaluated by constructing estimators from the Kubo integral \cite{sandvik_generalization_1992}
\begin{equation}
 \label{eq:Kubo}
 \chi_{AB}=\int_0^\beta\,\mathrm{d}\tau\left<A(\tau)B(0)\right>
\end{equation}
where the integrand shows the ensemble average of an imaginary-time dependent product with 
operators $A(\tau)=\mathrm{e}^{\tau H}\,A(0)\,\mathrm{e}^{-\tau H}$. For the case of diagonal 
operators $A$ and $B$ with eigenvalues $a(k)$ and $b(k)$, respectively, 
this integral can simply be written by including eigenvalues from all the propagated 
states \cite{sandvik_generalization_1992,sandvik_finite-size_1997},
\begin{equation}
 \label{eq:estimator}
 \chi_{AB}=\left<\frac\beta{n(n+1)}\left[\left(\sum_{k=0}^{n-1}a(k)
 \right)\left(\sum_{k=0}^{n-1}b(k)
 \right)+\sum_{k=0}^{n-1}a(k)b(k)
 \right]
 \right>.
\end{equation} 
For the conserved quantity magnetization $\mathcal M$, Eq.~(\ref{eq:estimator}) reduces to 
the uniform susceptibility $\chi_u$ with $a(k)=b(k)=\mathcal M$,
\begin{equation}
 \label{eq:ususc}
 \chi_u=\beta\left<\mathcal M^2\right>
\end{equation}
and for the quantity staggered magnetization $\mathcal M_s$, Eq.~(\ref{eq:estimator}) gives 
the staggered susceptibility $\chi_s$ with $a(k)=b(k)=\mathcal M_s(k)$,
\begin{equation}
 \label{eq:asusc}
 \chi_s=\left<\frac\beta{n(n+1)}\left[\left(\sum_{k=0}^{n-1}\mathcal M_s(k)
 \right)^2+\sum_{k=0}^{n-1}\mathcal M_s^2(k)
 \right]
 \right>.
\end{equation}
The staggered structure factor can be extracted from the second part of the Eq.~(\ref{eq:asusc}) in 
runtime, which can be defined as follows,
\begin{equation}
 \label{eq:structure}
 \mathcal S(\pi,\pi)=N\left<\mathcal M_s^2\right>.
\end{equation}
For each disorder strength $\rho=0.0,0.1,0.2,\cdots, 0.9,1.0$ the relevant quantities have been calculated 
for temperature values up to $T/J=2$.  $1000$ random 
realizations of the system have been generated for each disorder parameter to get a satisfactory statistics, 
and each average has been used as a bin which 
consists of at least $5\times10^5$ Monte Carlo steps (MCS) after discarding $5\times10^4$ MCS for the data 
analysis. To monitor the sample-to-sample fluctuations the running averages of the uniform susceptibility and 
the specific heat have been calculated in the vicinity of broad maximums and crossing points. Based on this, it 
is possible to say that 1000 independent realizations are found to be enough for good statistics. The 
standard errors have been propagated with the Bootstrap resampling technique for nonlinear functions. 
The spin gap values have been calculated by linearizing the Eq.~(\ref{eq:susc}) and making a least square fit to it 
at low temperatures to find the parameter $\Delta$.

\section{\label{sec:sec3}Results and Discussion}
The temperature dependencies of the calculated quantities are around the 
clean case ($\rho=0.0$) for all the disorder parameters of the system. While the disorder in the spin-spin couplings
does not cause a change in the physics of the results, the considered system has a special 
temperature point at which the specific heat take the same value regardless of the 
strength of the disorder. The same finding is also observed for the uniform susceptibility with a small difference in the location of the special point. A fine sweeping around these special points has been performed to validate the existence of this coincidence of the relevant curves. 

\begin{figure}[ht!]
\includegraphics[scale=0.425]{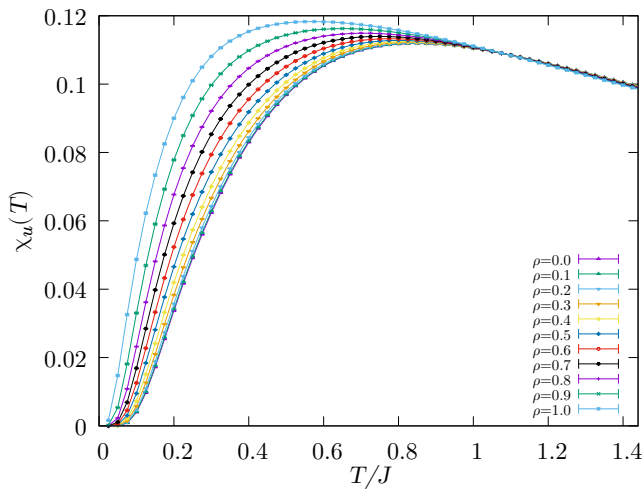}%
\caption{\label{ususc}(Color online) Thermal variation of the uniform susceptibility for the clean case and varying values of the 
disorder ratios: $\rho=0.0, 0.1, 0.2$,...,$0.8, 0.9$ and $1.0$. The lines are added  to guide the eye.}
\end{figure}

Thermal variation of the uniform susceptibility for several disorder strengths are displayed in Fig.~\ref{ususc}.
It is clear from the figure that the uniform susceptibility is nearly independent of the value of the disorder parameter at high temperature region and a Curie behavior is present in the system. The broad maximum of the uniform susceptibility shifts to the left with a slight increment in its value as the disorder parameter takes larger values. Also, an exponentially decreasing behavior is present at low temperature region for all disorder parameters which indicates the existence of a spin gap. 

\begin{figure}[ht!]
\includegraphics[scale=0.425]{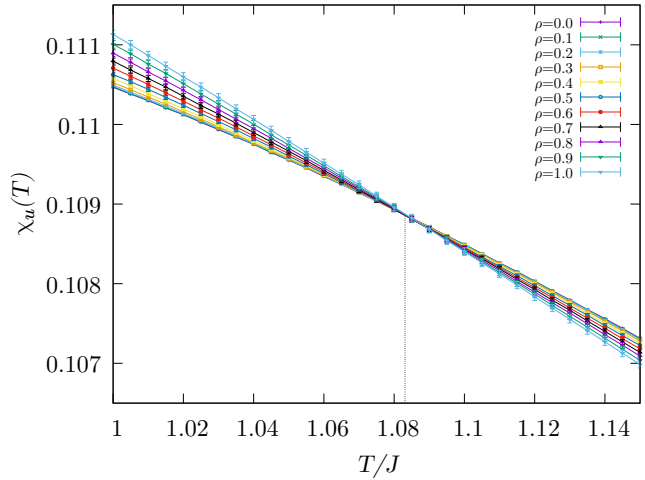}%
\caption{\label{ususcFine}(Color online) Fine sweeping of the uniform susceptibility near the crossing point. 
The vertical dashed line corresponds  to the calculated value of crossing point of the uniform 
susceptibility: $1.083$. The lines are added  to guide the eye.}
\end{figure}

We have calculated the crossing point for the uniform susceptibility using
the intersections of the following pairs $(\rho, \rho+0.3)$ of the disorder strengths: $(0, 0.3)$, $(0.1, 0.4)$, 
$(0.2, 0.5)$, $(0.3, 0.6)$, $(0.4, 0.7)$, $(0.5, 0.8)$, $(0.6, 0.9)$ and $(0.7, 1.0)$. We should also note that a 
number of $10^6$ MCS have been used for each configuration. As shown in Fig.~\ref{ususcFine}, our numerical findings 
suggest that the crossing temperature point is $1.083(3)$ for the uniform susceptibility.  Using the same protocol, we have also 
estimated  the corresponding uniform susceptibility value at the relevant crossing point to 
be $\chi^*_u=0.1088(1)$.  A similar point has been reported for magnetic spin susceptibilities in 
the spin-$1/2$ stacked 2-leg ladder systems \cite{johnston_magnetic_2000}. 
 
\begin{figure}[ht!]
\includegraphics[scale=0.425]{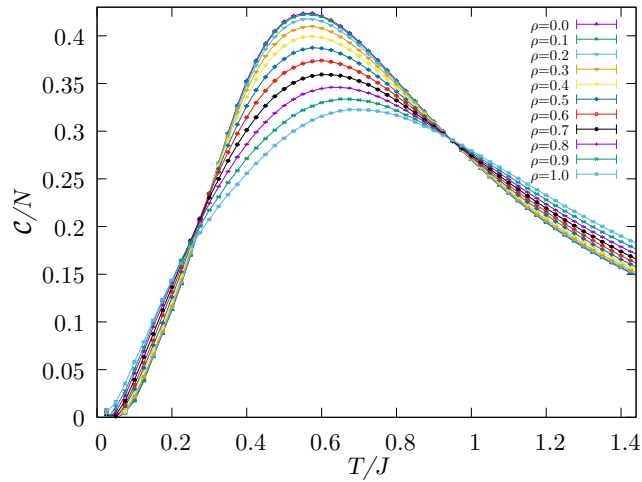}%
\caption{\label{specHeat}(Color online) Thermal variation of the specific heat curve for the clean case and  all disorder 
ratios:  $\rho=0.0, 0.1, 0.2$,...,$0.8, 0.9$ and $1.0$.  The lines are added  to guide the eye.}
\end{figure}

\begin{figure}[ht!]
\includegraphics[scale=0.425]{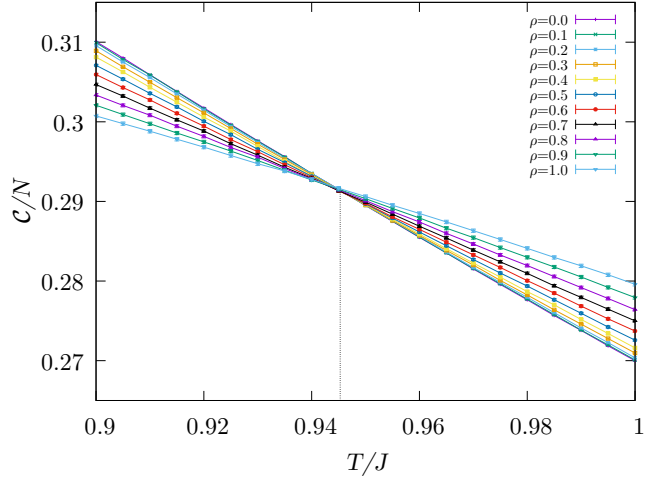}%
\caption{\label{specHeatFine}(Color online) Fine sweeping of the specific heat curves in the vicinity of the 
crossing point. The vertical dashed line corresponds  to the calculated value of 
crossing point of the specific heat: $0.9453$. The lines are added  to guide the eye.}
\end{figure}

As opposed to the uniform susceptibility, the maximum values of the specific heat tend to decrease with 
increasing value of disorder parameter as shown in the Fig.~\ref{specHeat} and the special temperature 
point is more visible. Fig.~\ref{specHeatFine} shows the fine sweeping around the special temperature point. 
By benefiting from the pairs of disorder strength mentioned for the uniform susceptibility, the crossing point is estimated as $0.9453(3)$ with a corresponding specific heat value of $\mathcal C^*=0.2912(7)$. These special points, for the specific heat and the uniform susceptibility, suggest that the point where all curves intersect can show a small difference depending on the quantity to be measured. Simulations with different system sizes up to $L_x=512$ have shown that the crossing points are almost size independent, which leads to negligible variations in their values. Based on this finding, it is possible to say that there are two distinct  
crossing points in the system.  Crossing points for the specific heat have been reported in various systems 
experimentally \cite{vollhardt_characteristic_1997,chandra_nearly_1999} and numerically \cite{georges_physical_1993}.
The special point is found to be independent of the parameters such as pressure, magnetic field and the local interaction 
of the Hubbard model. A theoretical origin of the special point has been investigated for lattice models and 
continuum systems \cite{vollhardt_characteristic_1997} and the numerical results have been given for the half-filled Hubbard model in all dimensions \cite{chandra_nearly_1999}. According to these studies, the specific  heat 
values are nearly the same  despite the corresponding crossing temperatures are different from each 
other for all dimensions. A nearly universal crossing value of the specific heat is obtained as $\approx0.34/k_B$, 
which is a little bit higher than that of obtained for the rung disordered Heisenberg ladder model considered here, i.e., $\approx0.29/k_B$. 
As in the case of Ref.~\onlinecite{vollhardt_characteristic_1997}, it should be noted that  the rate of 
change of specific heat values with respect to the disorder parameter changes its sign at the 
crossing point to make the total entropy change to zero for the present system. Furthermore, the crossing 
point of the specific heat can be considered as an inflection point.

\begin{figure}[ht!]
\includegraphics[scale=0.425]{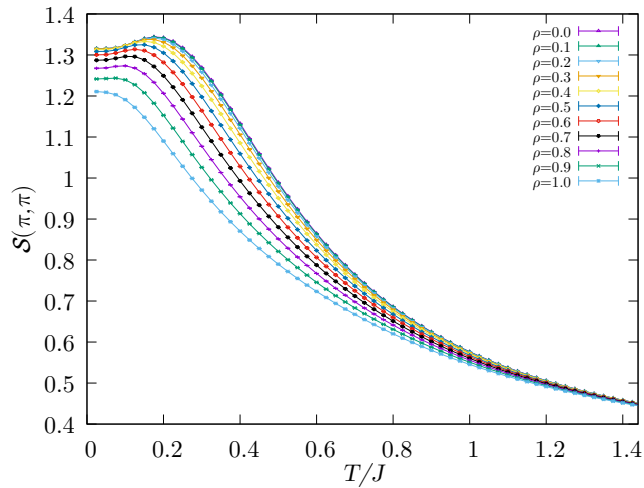}%
\caption{\label{structure}(Color online) Thermal variation of the structure factor for the clean case and varying values of  disorder ratios: 
$\rho=0.0, 0.1, 0.2$,...,$0.8, 0.9$ and $1.0$.  The lines are added  to guide the eye.}
\end{figure}

For even-leg ladders, the structure factor has a peak at a temperature that is below the relevant 
spin gap \cite{greven_monte_1996}. As it is shown in Fig.~\ref{structure} for this system the peaks shift to 
lower temperature region and decrease with an increment in disorder parameter. This shows the evidence of a decreasing spin gap with 
increasing disorder parameter value and this observation is also confirmed by calculating the value of the spin gap 
using Eq.~(\ref{eq:susc}). For high temperature values, the numerical results seem to be independent of the disorder parameter strength. On the other side, no crossing point is monitored in the temperature interval considered in the present study.

\begin{figure}[ht!]
\includegraphics[scale=0.425]{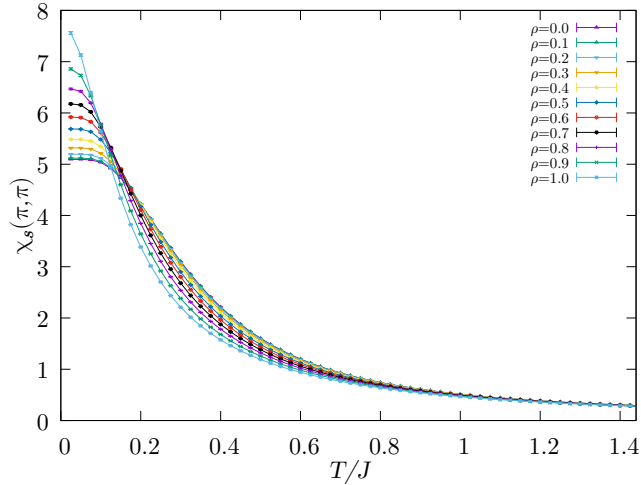}%
\caption{\label{staggered_susc}(Color online) Thermal variation of the staggered susceptibility for the clean case and 
all considered disorder strengths: $\rho=0.0, 0.1, 0.2$,...,$0.8, 0.9$ and $1.0$. The lines are added  to guide the eye.}
\end{figure}

The staggered susceptibility has a finite value at zero temperature for all disorder parameters as can be seen 
from Fig.~\ref{staggered_susc}. It also means that adding quenched disorder does not affect the ground state property of the system which is known to be close to the rung-dimer state in the clean case \cite{miyazaki_susceptibilities_1997}. Our QMC simulation results show that the obtained results are almost independent of the disorder parameter value at higher temperature region and no crossing point emerges for the staggered susceptibility.

\begin{figure}[ht!]
\includegraphics[scale=.4]{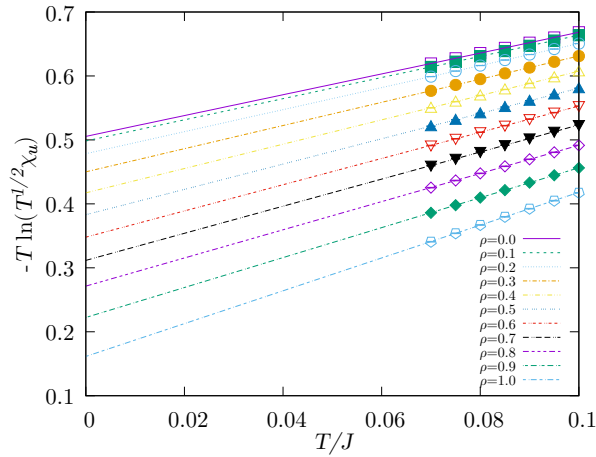}%
\caption{\label{gap}(Color online) Spin gap fit lines for the clean case and  all disorder ratios: $\rho=0.0, 0.1, 0.2$,...,$0.8, 0.9$ and $1.0$.  
The data point at $T/J=0.1$ for $\rho=1.0$ has been excluded from the fitting.}
\end{figure}

\begin{figure}[ht!]
\includegraphics[scale=.4]{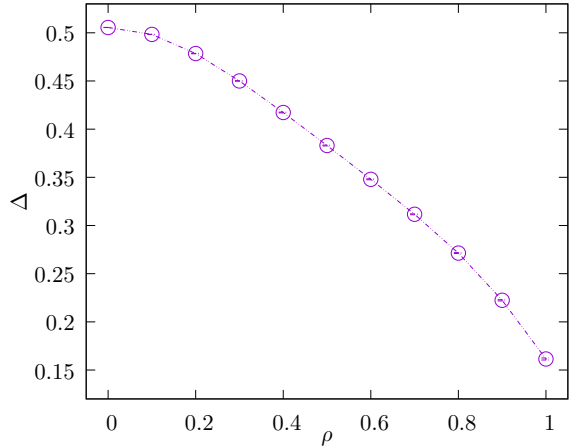}%
\caption{\label{dis_gap}(Color online) The disorder ratio dependence of the the spin gap value. The lines are added to guide 
the eye.}
\end{figure}

As depicted in Fig.~\ref{gap}, calculated spin gap values of the system are 
below the spin gap value of the clean case for all disorder parameters. It is also found that the spin gap values 
decrease with increasing disorder strength, leading an increment in the slopes of the relevant lines. 
For the limiting disorder parameter $\rho=1.0$, the spin gap is around $\Delta/J\approxeq0.16$ which 
is well above the weak coupling limit \cite{greven_monte_1996}. The value of the spin gap does not noticeably deviate 
from the clean case for $\rho=0.1$. As a final investigation, the variation of the spin gap with disorder coupling ratio is 
given in Fig.~\ref{dis_gap}. The decrement in the spin gap is nearly linear with the disorder parameter in the intermediate region. In particular, the spin gap declines slowly near the clean case and rapidly near to the fully disordered case.

\section{\label{sec:sec4}Conclusions}
In the present paper, we used the SSE QMC technique to study the thermodynamic properties of a two-leg ladder system with the quenched random bond 
disorder only among  the rungs of the ladder. Our simulation results show that there is a special point character 
in the system where the numerical results are independent of the disorder strengths for the specific heat and 
the uniform susceptibility, separately. The numerical values of 
these special points may be considered as a (nearly) universal value for the spin ladder systems. The 
numerical outcomes reported here also show that the averages of the disordered configurations 
do not tend to exhibit so different properties from the pure part.  This may be a result of the bond randomness 
including the same kind of interactions, which is introduced only in the rung direction of the ladder system. 
Another important result emerging in this study is that the spin gap values are found to 
decrease with increasing disorder parameter, as in the case of decreasing rung coupling values 
in the clean system. Finally, it would 
be interesting to study systems with disorder along only the leg or in both directions, 
as the disorder effects may exhibit interesting physical properties. Such kind of 
study may be the subject of future work. On the theoretical side, the equivalence of 
the half-filled Hubbard and the Heisenberg models might lead to exact expressions to 
extract crossing points.
 
\begin{acknowledgments}

The authors would like to thank S. Wessel for many useful comments and discussion on the manuscript. 
The numerical calculations reported in this paper were performed at T\"{U}B\.{I}TAK ULAKB\.{I}M
(Turkish agency), High Performance and Grid Computing Center (TRUBA Resources).
\end{acknowledgments}

\bibliography{apstemplate}
\end{document}